\documentclass{emulateapj}
\usepackage{amsmath}
\usepackage{booktabs}
\usepackage[stretch=10]{microtype}
\begin{document}

\title{The Polarization Signature of Local Bulk Flows}

\author{Elinore Roebber\altaffilmark{$\star$} \& Gilbert Holder}
\affil{Department of Physics, McGill University, Montr\'eal QC, H3A 2T8, Canada}
\altaffiltext{$\star$}{Email: roebbere@physics.mcgill.ca}

\shorttitle{Polarization Signature of Local Bulk Flows}
\shortauthors{Elinore Roebber \& Gilbert Holder}

\begin{abstract}
A large peculiar velocity of the intergalactic medium produces a Doppler shift of the 
cosmic microwave background with a frequency-dependent quadrupole term.  This 
quadrupole will act as a source for polarization of the cosmic microwave background, 
creating a large-scale polarization anisotropy if the bulk flow is local and coherent on 
large scales. In the case where we are near the center of the moving region, the 
polarization signal is a pure quadrupole. We show that the signal is small, but
detectable with future experiments for bulk flows as large as some recent
reports.
\end{abstract}


\section{Introduction}

The $\Lambda$CDM model of structure formation has been enormously 
successful in explaining a wealth of cosmological observations, including the 
cosmic microwave background \citep{bennett13, planck1_13},
galaxy clustering \citep{sanchez12}, and the brightness of distant supernovae 
\citep{conley11}.
However, it remains an active area of research, and there have been some 
peculiar observations: there may be some statistically unlikely features in the 
cosmic microwave background (CMB) fluctuations \citep{bennett11, planck23_13}, 
and there are hints that the local velocity field is not straightforward to 
understand in the $\Lambda$CDM framework 
\citep{watkins09, lavaux10, feldman10, kashlinsky10, kashlinsky12, colin11, ma11, 
magoulas13}, although many studies see good consistency
\citep{keisler09, dai11, osborne11, nusser11, turnbull12, mody12, branchini12,
ma13, rathaus13}.

Independent probes of such anomalies would be invaluable. For example, 
\citet{zhang11} used small-scale temperature anisotropies in the CMB as a 
probe of large-scale inhomogeneities in the universe. 

In this work we demonstrate that large angle CMB polarization can be similarly 
used to measure large-scale bulk flows.  In \S 2 we show this polarization signal, 
along with its unique frequency dependence, \S 3 estimates the expected
magnitude of the effect, \S4 uses data from the Wilkinson Microwave Anisotropy 
Probe (WMAP) to put limits on this signal, while \S 5 forecasts future constraints 
possible on large-scale bulk flows.

\section{CMB Anisotropy from a Local Bulk Flow}

CMB anisotropy caused by our peculiar velocity is a well-understood phenomenon, 
as the dipole has long been used as a measurement of the bulk velocity of the 
Earth \citep[e.g.][]{smoot77} and the annual modulation of the dipole due to the 
Earth's orbit is now used as a CMB calibration source \citep{bennett13, planck6_13}.
\citet{planck27_13} has also measured the effect on the small-scale CMB 
anisotropies due to the Doppler `modulation' and `aberration' resulting from our 
motion relative to the CMB.

In a related effect, electrons with peculiar velocities act as a source for CMB 
polarization \citep{baumann03}.  An electron moving with peculiar velocity 
$\beta \equiv v/c$ sees a boosted CMB spectrum:
\begin{equation}
	I_\nu = Cx^3 \left[\exp\left(x \frac{1+\beta\mu}{\sqrt{1-\beta^2}} \right) -1\right]^{-1}
\end{equation}
where $C=2(k_B T_\text{CMB})^3/(hc)^2$,   $x= h\nu/k_B T_{\text{CMB}}$ and $\mu$ 
is the cosine of the angle between the incoming photon direction and the direction of 
the velocity.  Expanding in terms of $\beta$ we see that
\begin{align} 
	I_\nu = 	& \frac{Cx^3}{e^x-1} \left \{ 1- \frac{xe^x}{e^x - 1} \mu \beta  \right.	\nonumber	\\
			&	\left. {} + \frac{xe^x}{e^x - 1} \left[  \left( \frac{x}{2}\coth \frac{x}{2}\right) \mu^2 + 
				\frac{1}{2} \right] \beta^2 +  \mathcal{O}(\beta^3) \right \}.
	\label{Ibeta}
\end{align}

The first order term in $\beta$ is the well-studied CMB dipole, with the frequency 
dependence of a pure fluctuation in temperature:  $(Cx^3/e^x-1)(xe^x/e^x - 1)$.
At second order, there is a component that again corresponds to a pure
temperature fluctuation, but also a term with a non-thermal spectrum,
which has been discussed extensively in \citet{kamionkowski03}.  

\begin{figure}
	\centering
	\includegraphics[width=0.48\textwidth]{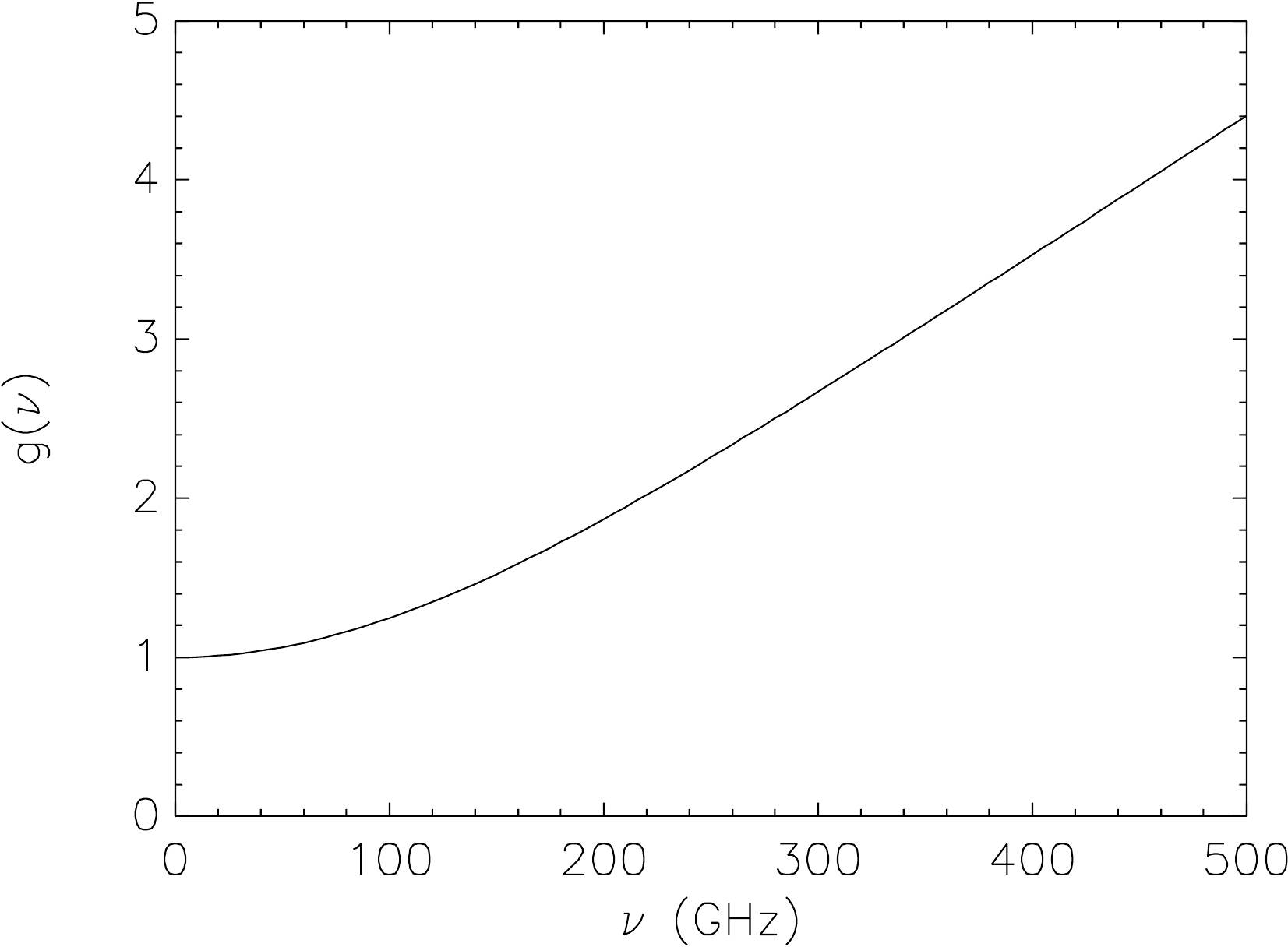}
	\caption{The non-thermal frequency dependence of the quadrupole anisotropy 
		induced by a peculiar velocity. }
	\label{fig:g} 
\end{figure}

This non-thermal second-order term can be decomposed into quadrupole 
($\propto \mu^2-1/3$) and monopole terms, so that if we divide out the 
thermal frequency dependence and expand it in spherical harmonics, it becomes
\begin{align}
	\left(\frac{x}{2}\coth \frac{x}{2}\right) \mu^2  =   
	     2\sqrt{\frac{4\pi}{5}}  \left( \frac{x}{2}\coth \frac{x}{2}\right) \left(\frac{1}{3} Y_{20} +  
	     	\sqrt{4\pi} Y_{00}\right). 
\end{align}
The coordinate system has been set by aligning the z-axis with the direction of 
motion.  In temperature units, the `kinematic' quadrupole seen by the electron 
and resulting from its motion is therefore
\begin{equation}
	a^{e^-}_{20} = \frac{2}{3} \sqrt{ \frac{4\pi}{5} } g(x) T_\text{CMB} \beta^2, 
\end{equation}
where the non-thermal frequency dependence is given by 
$g(x) = x/2 \, \coth (x/2)$ (see Fig.~\ref{fig:g}). 

More generally, if the velocity points in the direction 
$\mathbf{\hat{v}} = (\theta_v, \phi_v)$ and the incoming photon direction is 
$\mathbf{\hat{n}}$, the quadrupole term in \eqref{Ibeta} becomes
\begin{align}
	  \mu^2 - \frac{1}{3} 	& = \frac{2}{3} P_2(\mathbf{\hat{v}} \cdot \mathbf{\hat{n}})			\\
	  				& = \frac{8\pi}{15} \sum_{m=-2}^2 Y_{2m}^* (\theta_v, \phi_v) Y_{2m}(\mathbf{\hat{n}}) 
\end{align}
where $P_2$ is the second order Legendre polynomial.  Therefore, the 
kinematic quadrupole moments of the local radiation field can be expressed as 
\begin{equation}
	a^{e^-}_{2m}  = \frac{8\pi}{15} g(x) T_\text{CMB} \beta^2 Y^*_{2m} (\theta_v, \phi_v).
	\label{ae_2m}
\end{equation}

Thomson scattering of such a quadrupole anisotropy in the light
seen by the electron will create linear polarization in the scattered radiation.  
In particular, the $a^{e^-}_{2\pm 2}$ quadrupole will generate polarization 
along the outgoing $z$-axis~\citep{kosowsky99}. The Stokes $Q$ and $U$
parameters of the scattered radiation along the $\hat{z}$ axis are
\begin{equation}
Q(\hat{z}) \pm iU(\hat{z}) = \sigma_T \sqrt{\frac{3}{40\pi}} a^{e^-}_{2, \mp 2} \ .
\end{equation}

We will use this result to calculate the polarization emitted in all directions.  Following 
\citet{kosowsky99}, we rotate the the incoming field through the Euler angles 
$(\theta_e, \phi_e)$ so that the multipole coefficients in this rotated basis are
\begin{equation}
	\tilde{a}_{\ell m}^{e^-} = \!\!\! \sum_{m'=-m}^m
	\!\!\!  \mathcal{D}_{m' m} ^{\ell \, *}  (\theta_e, \phi_e, 0)  
	a^{e^-}_{\ell m'}
\end{equation}
where $ \mathcal{D}_{m^\prime m} ^{\ell }$ is the Wigner rotation symbol.  The 
polarization emitted in the direction $(\theta_e, \phi_e)$ will be generated by 
$\tilde{a}^{e^-}_{2, \pm 2}$:
\begin{equation} 
	\tilde{a}^{e^-}_{2,\pm2}  = \sum_m 	\mathcal{D}_{m, \pm 2} ^{2 \, *} (\theta_e, \phi_e, 0)  \, a_{2m}^{e^-}.
\end{equation}

The Wigner rotation can be expressed in terms of spin-weighted spherical 
harmonics~\citep{goldberg67, hu97}:
\begin{equation}
	\tilde{a}^{e^-}_{2,\pm 2} = \sqrt{ \frac{4\pi}{5} } \sum_m 
						{_{\mp 2}}Y_{2m}^* (\theta_e, \phi_e) \, a_{2m}^{e^-}.
\end{equation}

Therefore, the polarization emitted in the direction $(\theta_e, \phi_e)$ from scattering 
by a single electron moving in 
the direction $(\theta_v, \phi_v)$ is
\begin{align}
	Q(\theta_e, \phi_e) \pm i U(\theta_e, \phi_e)  = \sigma_T \sqrt{ \frac{3}{50} } \sum_m \!  
										{_{\pm2}}Y_{2m}^* (\theta_e, \phi_e)  \, 
								   		a_{2m}^{e^-}.
\end{align}

\begin{figure}[b]
	\centering
	\includegraphics[width=0.4\textwidth]{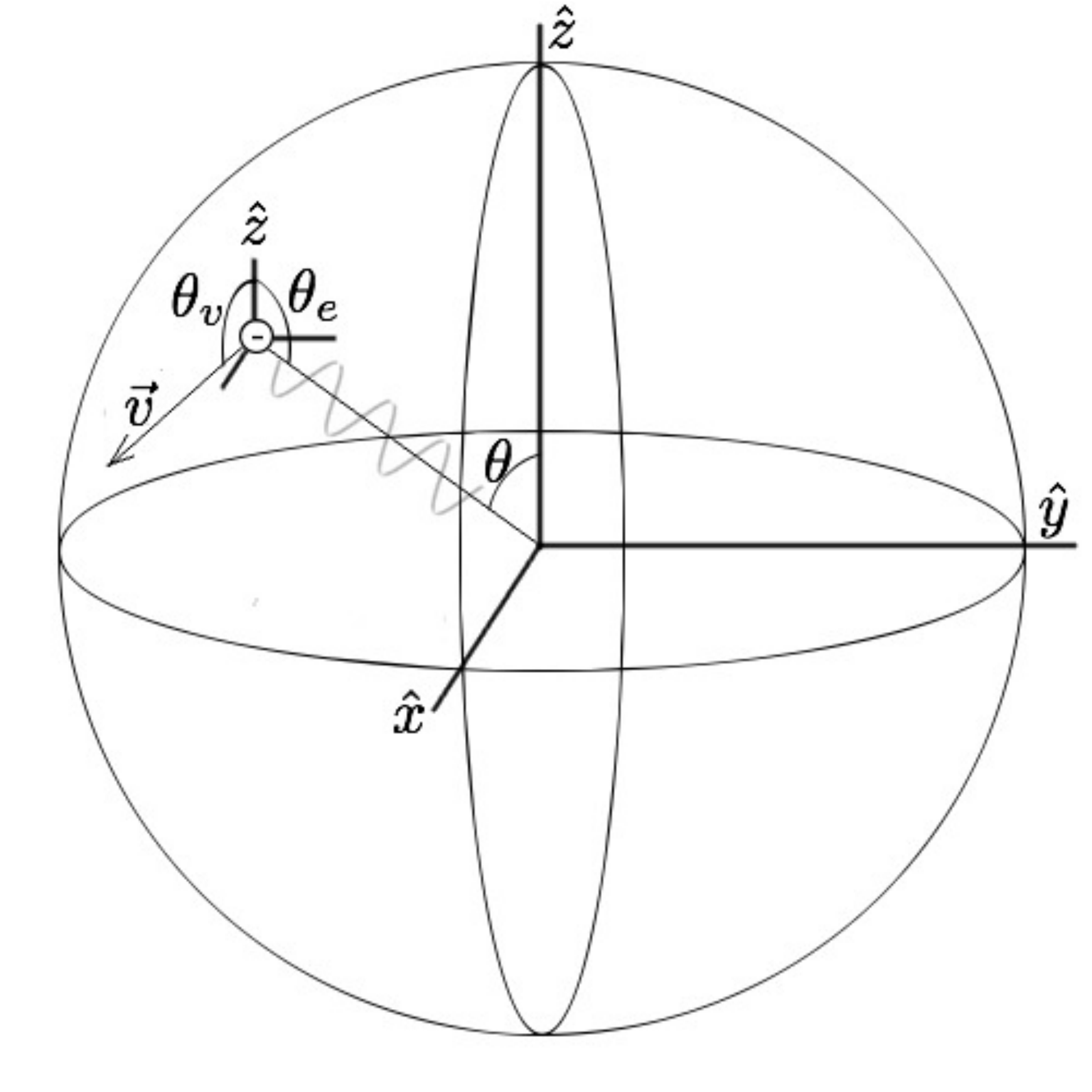}
	\caption{The geometry of a single polarized Thomson scattering as seen by an observer at the origin.}
	\label{fig:coords}
\end{figure}

To express this in terms of the polarization measured in direction $(\theta,\phi)$ by  
an observer (see Fig.~\ref{fig:coords}), note that $\theta_e = \pi - \theta$ and 
$\phi_e = \phi + \pi$.  We find that 
\begin{equation}	
	{_{\pm2}}Y^*_{2m} (\theta_e, \phi_e) = (-1)^{|m|} {_{\pm2}}Y_{2,-m} (\theta, \phi), 
\end{equation}
and so the polarization observed in the direction $(\theta, \phi)$ is 
\begin{equation}
	Q(\theta, \phi) \pm i U(\theta, \phi) = \sigma_T \sqrt{ \frac{3}{50} } 
		\sum_m  (-1)^{|m|}  {_{\pm 2}}Y_{2m} (\theta, \phi) \, a_{2, -m}^{e^-}.
	\label{QU_theta_phi}
\end{equation}

Any parcel of electrons that is moving relative to the CMB will 
therefore generate polarization anisotropy. If a large local volume has a 
substantial bulk flow there could be a large
scale polarization signal.  The total polarization can be calculated by substituting \eqref{ae_2m} into 
\eqref{QU_theta_phi} and 
integrating the contribution for all electrons along the line of sight: 
\begin{align} 
	 Q(\theta, \phi) \pm \,& i U(\theta, \phi)  =  g(x) T_{\text{CMB}}  \sqrt{ \frac{3}{50} } \frac{8\pi}{15}  \nonumber	\\
				& \times \int dr \Big( a(r) \sigma_T n_e(\theta, \phi, r) \beta^2(\theta, \phi, r)  \nonumber		 \\
				&  \times \sum_m   Y_{2m} [\theta_v(r,\theta,\phi), \phi_v(r,\theta,\phi)] \,  {_{\pm2}}Y_{2m} (\theta, \phi) \Big).
	\label{QUtotal}
\end{align}

For a general density and velocity distribution, equation
\eqref{QUtotal} leads to a complicated polarization pattern, as
there could be spatial (radial and angular) variation of $n_e$, 
$\beta$, $\theta_v$, and $\phi_v$.  

In cases where there is some symmetry,
it can be helpful to re-express \eqref{QUtotal} in terms of the E and B
modes defined by sums and differences of the spin-2
spherical harmonics \citep{zaldarriaga97}.
If there is no angular dependence to the electron densities and
velocities, the expansion coefficients for the spin-2 spherical harmonics
can be simply read off the above equation as the coefficients of ${_{\pm2}}Y_{2m} (\theta, \phi)$.  

In a model where the observer is located at the center of a spherical region of 
a bulk flow with constant direction but radially varying amplitude 
and constant electron density,  the signal will be a pure E-mode 
quadrupole:
\begin{align}
	a^E_{2m} 	& \equiv -\frac{1}{2} ({_2}a_{2m} + {_{-2}}a_{2m} )							\\
			& = -  g(x) T_\text{CMB} \frac{8\pi}{15} \sqrt{ \frac{3}{50} }  \nonumber 			\\
			& \qquad \times \int dr \, a(r) \sigma_T n_e \beta^2(r)Y_{2m}(\theta_v(r), \phi_v(r))
	\label{aE}
\end{align}
and
\begin{align}
	a^B_{2m}	& \equiv - \frac{i}{2} ({_2}a_{2m} - {_{-2}}a_{2m} ) = 0	
\end{align}

In the even simpler case where we assume a sphere of constant velocity out to a redshift $z_r$, 
the integral $\int dr \, \sigma_T n_e a(r)$ is equivalent to the optical depth to $z_r$, $\tau(z_r)$, 
and \eqref{aE} becomes 
\begin{equation}
	a^E_{2m}   = -  g(x) T_\text{CMB} \tau(z_r) \beta^2  \frac{8\pi}{15} \sqrt{ \frac{3}{50} } Y_{2m}(\theta_v, \phi_v).
	\label{aE_ang}
\end{equation}
In a coordinate system where the z-axis is aligned with the
bulk flow direction, the entire signal is contained in 
a single quadrupole moment with magnitude
\begin{equation}
	a^E_{20} = -  \frac{2}{5} \sqrt{ \frac{2\pi}{15} } g(x) T_\text{CMB} \tau(z_r) \beta^2.
\end{equation}

Since the velocity of the bulk flow, $\beta^2$, and its extent, $\tau(z_r)$, 
are degenerate, we will reparameterize equation \eqref{aE_ang}, using 
\begin{equation}
A_\text{flow} = \tau(z_r) \beta^2 , 
\end{equation}
so that
\begin{equation}
a^E_{2m}= -A_{\text{flow}} g(x) T_{\text{CMB}}  {\frac{8\pi}{15}} 
\sqrt{ \frac{3}{50} } Y_{2m}(\theta_v, \phi_v) .  
\label{aE_flow}
\end{equation}

\section{Polarization signatures of recent bulk flow measurements}

In recent years, there have been many surveys of local peculiar velocities.  Some of these surveys have 
reported consistency with the predictions of $\Lambda$CDM
\citep[e.g.][]{nusser11,turnbull12, dai11,keisler09,osborne11}
However, there have also been several reports of anomalously large values
\citep[e.g.][]{watkins09, lavaux10, feldman10,colin11, kashlinsky10,kashlinsky12}. 

One such measurement of a bulk flow somewhat larger than that predicted by linear-theory $\Lambda$CDM is 
due to \citet{feldman10}, who compiled the results of most prior major peculiar velocity surveys.  
They measure a bulk flow of $416\pm78$~km/s using a $100h^{-1}$Mpc Gaussian window, which they find to be in 
tension with $\Lambda$CDM. 

 A second case is given by the measurements of \citet{kashlinsky10} using the
kSZ signal produced by hot gas in galaxy clusters.  They report a bulk flow of 1000~km/s on scales of 
$\gtrsim 550 h^{-1}$Mpc, although the significance of this measurement has been contested 
\citep{keisler09, osborne11, mody12, lavaux13,planck_intermediate13_13}. This measurement is inconsistent with $\Lambda$CDM.  
Additionally, the authors propose that the bulk flow measured could be a piece of a `dark flow' 
extending out to the horizon, as would be expected for a `tilted universe' \citep{turner91, ma11}.

Since the reported bulk flows span a wide range of sizes, we investigate several scenarios to determine the 
possible resulting polarization signals.  Our four scenarios are summarized in Table~\ref{tab:signals}, and represent: 
\begin{itemize}
	\item a `small' bulk flow consistent with $\Lambda$CDM predictions
	\item a `medium' bulk flow in tension with $\Lambda$CDM, approximating the results of \citet{feldman10}
	\item a `large' bulk flow inconsistent with $\Lambda$CDM, approximating the results of \citet{kashlinsky10}
	\item a large `dark flow' similar to that proposed by \citet{kashlinsky10}, extending to at
		least the redshift of reionization (but not to recombination)
\end{itemize}

\begin{table}
\begin{tabular} {l c c c c} \toprule
	& Velocity (km/s) & $z_r$ & $-a_{20}^E$ (nK) & $A_\text{flow}$\\
	\midrule
	Small & 100 &0.03 & 0.006 & $6 \times 10^{-12}$ \\
	Medium & 400 & 0.03 & 0.1 & $9 \times 10^{-11}$ \\
	Large & 1000 & 0.25 & 6 & $6 \times 10^{-9}$ \\
	Dark Flow & 1000 & $\gtrsim 10.6$ & 1000 & $1\times10^{-6}$ \\
	\bottomrule
\end{tabular}
\caption{Polarization signal for local bulk flows at $\nu=150$~GHz}
\label{tab:signals}
\end{table}
	
For comparison, the cosmological E-modes on these scales are of order
100 nK on horizon scales, while B-modes from inflation on the largest 
scales would have an amplitude of order 1 nK for a tensor-to-scalar ratio of 
order $r\sim 10^{-4}$.   While it is a small signal, the characteristic frequency 
dependence $g(x)$ of the contribution from bulk flows makes such a 
measurement at least possible in principle.  

In comparison to the secondary anisotropies induced 
by the thermal Sunyaev-Zel'dovich effect (tSZ), with a nominal expected 
$y \sim 10^{-6}$, the bulk flow signal, which varies as $\tau \beta^2$, is 
smaller (see the $A_\text{flow}$ column in Table~\ref{tab:signals}).  
Indeed, its frequency dependence is a linear combination of the
frequency dependences of thermal fluctuations and tSZ.  However, it is 
a large-scale polarization signal, whereas $y$ fluctuations occur on 
small scales and are unpolarized to leading order.  We will discuss the 
possibility of confusion from the polarized tSZ in \S4.

\section{Constraints on bulk flows from large-angle CMB polarization data}

A very large bulk velocity extending across a significant portion of the 
universe  could be detectable by WMAP.  We will therefore use WMAP9 
polarization data \citep{bennett13} to put an upper limit on the bulk flow.  
We perform a $\chi^2$ fit of the WMAP data for $A_\text{flow}$, using the 
foreground-reduced V-band maps.  We extract the quadrupoles from the 
WMAP maps and compare them to equation \eqref{aE_flow} with freely 
varying $A_\text{flow}$ but velocity fixed in the direction found by 
\citet{kashlinsky10}: $(l,b)=(288^\circ,30^\circ)$.  We choose this direction 
in the hope of constraining their proposed `dark flow' since all other claimed 
signals would be too small to be measured by WMAP.  

The variance in the $\chi^2$ fit is given by 
\begin{equation}
	C_2 = C_2^\text{theory} +N_2,
\end{equation}
where $C_2^\text{theory}$ is the theoretical value of the EE power spectrum at 
$\ell=2$ and $N_2$ is the experimental noise of the E mode quadrupole, which we 
estimated by averaging the nearby WMAP BB $C_\ell$'s in the same map.
Since WMAP does not detect a physical B mode signal, the measured 
BB power spectrum is an estimate of the E mode 
noise on the same scales.

\begin{figure}
	\plotone{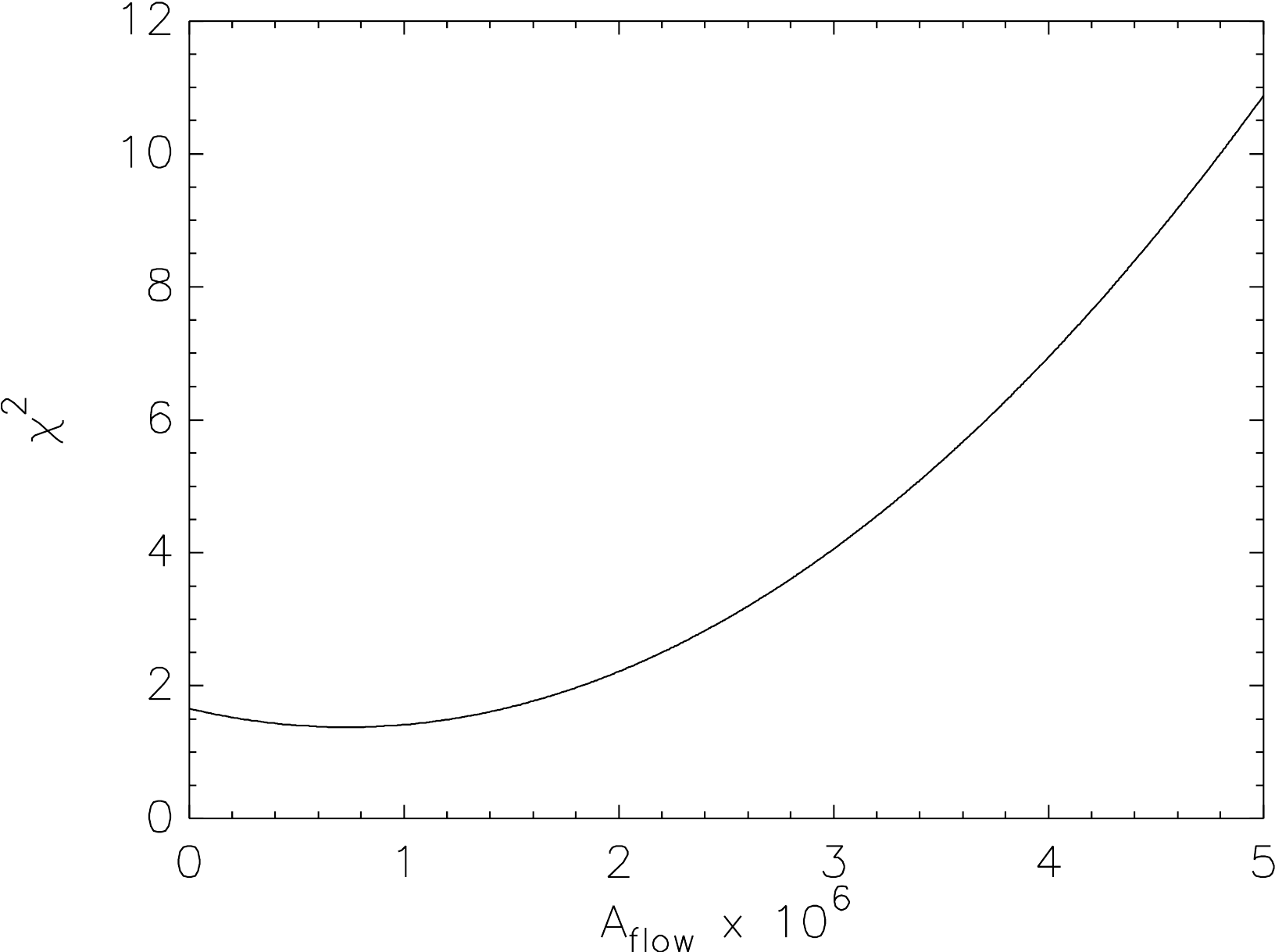}
	\caption{The results of the $\chi^2$ fit for $A_\text{flow}$ along the direction found by \citet{kashlinsky10}. } 
	\label{fig:chi2}
\end{figure}

Since the bulk flow signal is frequency-dependent, the foreground removal 
process may decrease the signal present in these maps.  However, $g(x)$ 
does not vary greatly over the WMAP bands, so we expect this to be a small 
effect. Constraints on the bulk flow amplitude are shown in Figure \ref{fig:chi2}.  
We find a 95\% upper limit of $A_\text{flow}=3\times 10^{-6}$, which
corresponds to a bulk flow of order 2000 km/s extending out to reionization. 
Stronger constraints will require lower noise, as well as multi-frequency 
combinations (to suppress primary CMB fluctuations).

With ongoing rapid improvements in polarization sensitivity, we 
can expect big improvements in these sorts of measurements.
At $\ell=2$, the expected polarization sensitivity of the Planck
experiment is of order 15 nK, only slightly greater than the signal  
expected for the large flow observed by \citet{kashlinsky10}.  
Future CMB satellites such as PIXIE \citep{kogut11} or EPIC 
\citep{bock08} could have sensitivity better than 2 nK, which 
would be sufficient to detect this signal.  Moreover, an experiment 
like PIXIE, which would make measurements in many frequency 
bands over a very wide range, would be ideal for separating 
the bulk flow signal from the primary CMB anisotropies.  

No experiments currently planned would be capable of measuring 
the signal predicted by $\Lambda$CDM, which are at least an 
order of magnitude lower in amplitude.  In general, the signal
expected from $\Lambda$CDM will look quite different from our model, 
since we have only included the the contribution of a local bubble.  
It is also small enough that there will be a term of comparable size from the 
polarized tSZ.  This signal results from the isotropic thermal motion of 
electrons, coupled to anisotropy in the CMB.  Since the frequency 
dependence of the bulk flow signal is a linear combination of the 
frequency dependence for a thermal fluctuation and a tSZ fluctuation, 
the polarized tSZ will ultimately place a limit on how precisely 
the bulk flow can be measured.

We can estimate the magnitudes of the two signals by considering that 
the bulk flow signal varies as $\tau\beta_\text{BF}^2 T_0$ (where $T_0$ is 
the CMB monopole), while the polarized tSZ goes as $\tau\sigma_e^2 T_2$ 
(where $T_2$ is the CMB quadrupole and $\sigma_e^2$ is the rms velocity 
of the electrons).  The terms $\sigma_e^2$ and $\beta_\text{BF}^2$ are both 
set by the local gravitational potentials, but the thermal energy of the electrons 
is enhanced by a factor of $m_p/m_e\sim 2000$. Since $T_2 \sim 10^{-5} T_0$, 
we estimate the polarized tSZ to be smaller than the bulk flow signal by a factor 
of $\sim10^{-2}$.  In general, we expect that for a $\Lambda$CDM bulk flow,
the polarized tSZ signal will generally be smaller but non-negligible for 
precise measurements.

The quadrupolar polarized foregrounds are of order 1~$\mu$K, so 
considerable foreground reduction will be required.  Additionally, since the 
frequency dependence of this signal is equivalent to a linear combination of 
the primary CMB and tSZ signals, leakage of  tSZ temperature signal into the 
polarization will form a significant source of error for attempted detections.   
For many experiments, an important source of temperature-polarization 
leakage results from a slightly elliptic beam, which, when rotated, leads to a 
spurious coupling of the polarization with the local unpolarized quadrupole.  

We expect that future CMB polarization experiments designed to detect 
primordial B modes may be able to measure the polarization produced by 
bulk flows.  For these experiments, the beam ellipticity should be of order 
$0.01$\% \citep{bock06}.  Extrapolating from the Planck tSZ power spectrum 
\citep{planck21_13}, we estimate a quadrupole term of $\sim 0.03$~$\mu$K, 
implying a spurious polarization signal due to tSZ with amplitude 
$\sim 0.003$~nK.  This is close in size to our `small' model, and represents 
another potential difficulty for detection. However, future experiments  will be 
able to effectively constrain anomalous large scale bulk flows.

\section{Discussion}

We have presented a new probe of large scale bulk flows in the universe.
In general, the induced polarization anisotropy will have a well-defined
frequency dependence but would have complicated angular structure. In 
the simple case of large local bulk flows, WMAP data is able to rule
out flows that are only slightly larger than the largest ones proposed.
Planck will have the sensitivity to rule out horizon-scale bulk flows
and put strong limits on the proposed ``dark flow'' of \citet{kashlinsky10}.

With a unique frequency dependence, these observations should be
separable from both primary CMB fluctuations and from foregrounds. While
the $\ell=2$ mode (quadrupole) is the simplest to understand as a simple
large-scale local bulk flow, there will be interesting signatures on
a variety of angular scales. While such bulk flows can also be
mapped using the kinetic Sunyaev-Zel'dovich effect, the polarization
signature offers a complementary view of the large scale motions of
cosmic gas.  The kinetic polarization signal is most sensitive 
to velocities perpendicular to the line of sight, whereas the kinetic 
Sunyaev-Zel'dovich effect is sensitive to velocities projected along the 
line of sight, suggesting that measurements of the two effects can be 
combined to measure the three-dimensional velocity field. 

Such large scale bulk flows are unlikely to exist in the framework of the 
$\Lambda$CDM model of structure formation, and represent one of a 
number of possible anomalous deviations from $\Lambda$CDM.  
In spite of the many measurements of our local velocity field, 
it is not yet clear whether the local peculiar velocities are consistent with 
expectations.  In particular, although the CMB dipole is believed
to arise from a Doppler shift due to our velocity with respect to the CMB, 
measurements of local peculiar velocities have not been able to 
completely account for it \citep{kocevski06}.  

As a result, it is of interest to have multiple independent probes of the 
large-scale velocity field. Since the bulk flow polarization signal described 
here is most sensitive to large-scale flows, unlike most standard peculiar 
velocity measurements, it may have the potential to provide a consistency 
test for better understanding the CMB dipole and the local velocity field on 
large scales.  
\\

We thank Olivier Dor\'e for valuable discussions.
This work was supported by NSERC, CIfAR, and the Canada Research Chairs program.
This work was supported in part by the National Science Foundation under Grant 
No.\ PHYS-1066293 and the hospitality of the Aspen Center for Physics.
Some of the results in this paper have been derived using the 
HEALPix\footnote{http://healpix.sourceforge.net} \citep{gorski05} package.
We acknowledge the use of the Legacy Archive for Microwave 
Background Data Analysis (LAMBDA), part of the High Energy 
Astrophysics Science Archive Center (HEASARC). HEASARC/LAMBDA 
is a service of the Astrophysics Science Division at the NASA 
Goddard Space Flight Center.

\bibliography{pecvel}

\end{document}